\documentclass[12pt]{iopart}

\usepackage{iopams}
\usepackage{mathrsfs}

\newcommand{\im}{{\rm i}}

\newcommand{\eqtri}{\triangleq}
\newcommand{\ket}[1]{\left|{#1}\right\rangle}
\newcommand{\bra}[1]{\left\langle{#1}\right|}

\newcommand{\Dcirc}{\mathcal{D}^{\circ}}

\newcommand{\tag}[1]{\tau[{#1}]}
\newcommand{\ext}[1]{\chi[{#1}]}
\newcommand{\extag}[2]{\chi[{#1}]\tau[{#2}]}

\begin{document}

\title[Tagged vector space, Part II]{Tagged vector space, Part II: function space index and the implied functional integration measure}

\author{Filippus S. Roux}
\address{University of Kwazulu-Natal, Private Bag X54001, Durban 4000, South Africa}
\ead{rouxf@ukzn.ac.za}

\begin{abstract}
The definition of quantum states in terms of tagged vector spaces is generalized to incorporate the spatiotemporal and spin degrees of freedom. Considering a tagged vector space where the index space is a function space, representing the additional degrees of freedom, we obtained axioms for the tags that include a completeness condition expressed in terms of a functional integral with an abstract functional integration measure. Using these axioms, we derive a generating functional for the moments of this functional integration measure. These moments are then used to evaluate the functional integrals of Gaussian functionals, leading to expressions in accordance with those obtained as generalizations of equivalent integrals over a finite number of integration variables. For a Gaussian functional used as a probability distributions, we show that its moments, obtained with this functional integration measure, satisfy Carleman's condition, indicating that the measure is unique.
\end{abstract}

\section{Introduction}

When the number of degrees of freedom in a quantum system becomes infinite, the formalism in terms of which such a system is modelled often leads to functional integrals, where the integration is performed over a function space. Examples include, the path integral found in quantum field theory \cite{peskin,weinberg1,rivers}, the Wiener integral defined for stochastic processes \cite{wiener2,wiener3}, and the functional integral for quantum optics on a functional phase space \cite{fpsm}. Functional integrals are often not mathematically well defined, apart from a few cases such as the Wiener integral in the context of probability theory. When the path integral is defined with an imaginary time, the integrand is converted into a probability distribution, allowing the use of the Wiener measure \cite{bsimon,gj}.

In more general scenarios, direct attempts to define integration measures for functional integrals as generalizations of integration measures over finite dimensional spaces (such as the Lebesgue measure or the Haar measure \cite{maatteorie}) are unsuccessful. The infinite number of degrees of freedom tends to render Borel measures infinite. Moreover, while the function space is equivalent to the product space of a infinite number of real lines, the latter is known to be \textbf{not} locally compact, disallowing a Haar measure. Therefore, a suitable mathematical definition for a functional integral within a particular context requires a different approach \cite{daniell}.

Here, we address this problem in a specific context, using a \textbf{tagged vector space} \cite{deeleen} where the tags are indexed by the elements of a function space. Such a formulation can, for example, be used to consider the Wigner functionals that are defined on a functional phase space \cite{fpsm}. Before addressing functional integration, we develop this functional representation as a generalization of the one-dimensional case \cite{deeleen}. The axiomatic completeness condition then becomes an abstract functional integral. The axioms of such an abstract functional tagged vector space lead to a generating functional for the moments of the implied functional integration measure. It allows the explicit evaluation of functional integrals of Gaussian functionals.

A normalized Gaussian functional can represent a probability distribution on a function space. The question whether the moments of such a probability distribution can be used to reproduce the probability distribution falls under the \textbf{moment problem} \cite{lin}. In the current context, it is represented by a functional version of the Hamburger moment problem. The Carleman's condition \cite{momentprob} provides a test with which one can determine whether the moments can perform this task. We demonstrate that the moments obtained from a normalized Gaussian functional with the implied functional integration measure satisfies Carleman's condition.

\subsection{Infinities}

In our endeavour to develop a rigorous mathematical treatment of functional integration within the given context, we inevitably encounter the irreducible appearence of infinities. It could present an obstacle if not treated in a suitable manner. Therefore, an adequate formal mathematical definition of functional integration in the current context cannot succeed without some way to deal with infinities.

In mathematics, the appearence of an infinity is often undesirable, even when dealing with infinitely many degrees of freedom. A quantity is only considered to be well-defined if it is finite. The reason is that basic calculations involving infinities (such as $\infty/\infty=\ ?$) lead to ambiguities. These ambiguities arise due to \textbf{cardinal arithmetic}, in which all infinities of the same kind (cardinal numbers) are treated as being equal. (There are predominantly two different kinds of cardinal numbers: the \textbf{cardinality of countable infinity} and the \textbf{cardinality of uncountable infinity} also called the \textbf{cardinality of the continuum}. Certain calculations, such as exponentiation, applied to the first leads the second.) For example, if $\alpha$ is an infinity (a cardinal number), then $\alpha+1=\alpha$, in contrast to the situation in \textbf{ordinal arithmetic}, which applies to all finite numbers (ordinal numbers).

In physics, systems with infinitely many degrees of freedom often leads to such infinities in calculations. To deal with them the usual approach is to use some form of \textbf{regularization} \cite{iz} in which the quantities that become infinite in the full calculation are rendered finite, perhaps by reducing the number of degrees of freedom to be finite. The calculation is then performed in this finite form, after which the appropriate limit is applied to recover the result of the calculation without the imposed regularization. In cases where the result must be finite (as for physically observable quantities) all the would-be infinities are removed in the limit, rendering the result finite. (The situation in quantum field theory, as used in particle physics, is more complicated and needs a subsequent process of \textbf{renormalization} \cite{peskin} to render the results finite.) Unfortunately, there are many different ways to perform such a regularization procedure and they may affect the calculation in different ways.

Standing back to consider what actually happens when using a regularization procedure, we see that we have in essence convinced ourselves that cardinal numbers can be treated like ordinal numbers, provided that we keep careful track of how they are carried along during the calculation.

During the regularization procedure, cardinal numbers are represented as would-be infinites in terms of distinguished ordinal numbers.
When the limit is applied, any would-be infinities are removed either because the inverse of such a quantity becomes zero or because they cancel. While the former mechanism works for any kind of infinity, the latter requires that the would-be infinites are identically equal; representing exactly the same quantity. If we can somehow keep careful track of how cardinal numbers are distinguished during the calculation, we would be able to ensure that the necessary cancellations take place to render the result finite. In such a case, it would be essential that would-be infinities are carefully distinguished based on how they are produced during or even before the calculation. In other words, if a certain part of a calculation (like the trace over the identity on an infinite domain) produces an infinity, we can label that quantity by a special symbol, which we call a \textbf{cardinal tag}. Then we treat that quantity like one would treat an unknown variable for an ordinal number in the full calculation. (In effect, the cardinal tag does not take part in the calculation because it does not evaluate to a numerical value, but is instead just ``dragged along'' in the calculation.) When the full calculation leads to something that must come out to be finite (such as a measurable quantity), then all these cardinal tags must cancel (or their inverses be rendered zero). This approach replaces the need for a regularization procedure and it is unambiguous. However, the need for care when assigning cardinal tags to would-be infinities cannot be overemphasized.

The implication is that infinities are not treated as being equal as per cardinal arithmetic. Instead the cardinality of countable infinity and the cardinality of the continuum are treated as spaces to which the different infinities, labeled by different cardinal tags, belong. These different infinities are then treated in terms of the cardinal tags as if they obey ordinal arithmetic. When done correctly, such ordinal calculations lead to finite results (in the calculation of measurable quantities) thanks to the cancellation of these cardinal tags.

In our development of functional integration, we'll encounter two distinguished infinities, which are labelled by their respective cardinal tags. Since our calculations sometimes produce functionals that are not measurable quantities themselves, one of these cardinal tags will remain. However, when such a functional is used in a subsequent calculation of a measurable quantity, this cardinal tag is removed (cancels out).

\section{Incorporating all degrees of freedom}

Before addressing functional integration, we first discuss the function space over which such integrations are to be performed. The definition of this function space is determined by the context within which it serves its purpose.

In \cite{deeleen}, we introduced a way to formulate a quantum theory in terms of a tagged vector space. Here, we are interested in modelling quantum optics in all degrees of freedom. It can for example be modelled in terms of a functional phase space on which Wigner functionals represent quantum states and operators \cite{fpsm}. The foundation for this functional phase space is the quadrature eigenvectors serving as bases that include all the degrees of freedom. Viewed in terms of a tagged vector space, the quadrature bases are replaced by tags, but it requires that the index space, which is used to label the tags, is a function space. Such a scenario is inevitable when the degrees of freedom of the quantum systems become infinite as is often found in physical scenarios such as particle physics and quantum optics.

\subsection{Functional tagged vector spaces}

As explained in \cite{deeleen}, a \textbf{tagged vector space} consists of all linear combinations of a set of \textbf{tags} $\{\tau\}$ indexed by the elements of an \textbf{index space} $\mathcal{I}$, with coefficients taken from a \textbf{coefficient space} $\mathcal{C}$. It is accompanied by an \textbf{adjoint tagged vector space} consisting of all linear combinations of a set of \textbf{extractors} $\{\chi\}$ indexed by the elements of the same index space with coefficients taken from a coefficient space which is usually the same as that of the tagged vector space. The purpose of an extractor is to extract the coefficient of a specific tag from a linear combination of tags.

The development of tagged vector spaces in the context of the Dirac formalism in \cite{deeleen} only provides the formalism for the quantum (particle-number) degree of freedom. By contrast, a formalism that incorporates all the degrees of freedom (spin and spatiotemporal degrees of freedom) is necessarily defined in terms of a \textbf{function space}. To define such a formalism, we generalize the concept of a tagged vector space to become a \textbf{functional tagged vector space}, in the context of a functional phase space.

In the case we are interested in here, the real quadrature field variables, which are the coordinates of a functional phase space, represent two \textbf{mutually unbiased index spaces}. The functions that are the elements (quadrature field variables) of these index spaces parameterize the spin and spatiotemporal degrees of freedom. We can assume that these functions are real Schwartz functions, so that $\mathcal{I}\equiv S(\mathbb{R}^3)$, which is dense in $L^2(\mathbb{R}^3)$. The type of Schwartz functions also depends on the nature of the spin degrees of freedom. However, the spin does not introduce complications and can be specified as a generalization of the scalar case without spin degrees of freedom. In what follows we'll generally ignore the spin degrees of freedom.

Although one can use Dirac notation to represent the tags and their associated extractors, we retain the original notation for the tags and extractors, representing the tags as $\tag{q}$ and $\tag{p}$ for the two quadrature field variables, respectively, and their associated extractors are represented by $\ext{q}$ and $\ext{p}$, respectively. The square backets indicate that the tags and extractors are functionals of the indices $q$ and $p$, as elements of the function spaces serving as index spaces. The axioms that define the properties of the tags and extractors are then expressed in functional form as
\begin{itemize}
\item \textbf{orthogonality}:
\begin{equation}
\extag{q_1}{q_2} = \delta[q_1-q_2] ~~~~~ {\rm and} ~~~~~ \extag{p_1}{p_2} = (2\pi)^{\Omega}\delta[p_1-p_2] ,
\label{ortax}
\end{equation}
where $\delta[\cdot]$ represents a \textbf{Dirac delta functional} and $\Omega$ is a cardinal tag for the number of degrees of freedom of the index space,
\item \textbf{completeness}:
\begin{equation}
\int \tag{q}\ext{q}\ \mathcal{D}[q] = \mathbb{I} ~~~~~ {\rm and} ~~~~~ \int \tag{p}\ext{p}\ \Dcirc[p] = \mathbb{I} ,
\label{volax}
\end{equation}
where $\mathcal{D}[q]$ and $\Dcirc[p]$ are abstract \textbf{functional integration measures}, with the latter incorporating a factor of $1/(2\pi)$ for each degree of freedom, and $\mathbb{I}$ is the identity operator,
\item \textbf{unbiased}:
\begin{equation}
\extag{q}{p} = \exp(\im q\diamond p) ~~~~~ {\rm and} ~~~~~ \extag{p}{q} = \exp(-\im q\diamond p) ,
\label{mubax}
\end{equation}
where the $\diamond$-contraction is defined as
\begin{equation}
q\diamond p \eqtri \sum_s \int q_s(\mathbf{k})p_s(\mathbf{k})\ \frac{{\rm d}^3 k}{(2\pi)^3} .
\end{equation}
in terms of the field variables $q_s(\mathbf{k})$ and $p_s(\mathbf{k})$, as functions of the wave vector $\mathbf{k}$. For the general case, we also show a spin index $s$, although the definition of the Schwart space that we use here does not incorporate spin.
\end{itemize}
The right-hand sides in (\ref{ortax}) and (\ref{mubax}) are tempered distribution. Although expressed in their bare forms, these axioms lead to expressions of linear operators with these distributions acting as Schwartz kernels. Note that the axioms for the $p$-index space incorporate factors of $2\pi$ to anticipate the definition of the Fourier transform.

To define the functional tagged vector space and the adjoint functional tagged vector space we need to define the coefficient space. However, it begs the question of the definition of the functional integration, as discussed below. For the sake of context, we provide these definitions under the assumption of well-defined functional integrals.

The coefficient space $\mathcal{C}$ is a space of normalized \textbf{Schwartz functionals} that map the index space to $\mathbb{C}$,
\begin{equation}
S[S(\mathbb{R})] \supset \mathcal{C} \eqtri \left\{ F : S(\mathbb{R}^3) \rightarrow \mathbb{C} , \|F\|=1 \right\} .
\label{ftagmap}
\end{equation}
The norm $\|F\|$ is an Euclidean norm defined in terms of a functional integral.

Now, we can define the functional tagged vector space in terms of the $q$-tags as
\begin{equation}
\Upsilon = \left\{\ket{F} = \int \tag{q} F[q]\ \mathcal{D}[q] : q\in S(\mathbb{R}), F\in S[S(\mathbb{R})] , \|F\|=1 \right\} ,
\end{equation}
and the adjoint functional tagged vector space that accompanies it as
\begin{equation}
\Upsilon' = \left\{\bra{G} = \int G^*[q] \ext{q}\ \mathcal{D}[q] : q\in S(\mathbb{R}) , G^*\in S[S(\mathbb{R})] , \|G\|=1 \right\} .
\end{equation}

The equivalent definitions in terms of $p$ produce the same functional tagged vector space and adjoint functional tagged vector space, because the tags that are indexed by $q$ are unitarily related to those indexed by $p$ via the unbiased axiom (\ref{mubax}). In other words,
\begin{eqnarray}
\mathbb{I}\ket{F} & = \int \tag{p}\extag{p}{q} F[q]\ \Dcirc[p]\ \mathcal{D}[q] \nonumber \\
& = \int \tag{p} \exp(-\im q\diamond p) F[q]\ \mathcal{D}[q]\ \Dcirc[p] \nonumber \\
& = \int \tag{p} \tilde{F}[p]\ \Dcirc[p] ,
\end{eqnarray}
where
\begin{equation}
\tilde{F}[p] = \int \exp(-\im q\diamond p) F[q]\ \mathcal{D}[q]
\end{equation}
is the functional Fourier transform. As a result, the elements of the tagged vector space can be represented in terms of tags indexed by either $q$ or $p$ (or any other index unitarily related to $q$) where the coefficient function indexed by $p$ is the Fourier transform of the coefficient function indexed by $q$. Moreover, regardless of the (unitarily related) index used in its definition, the ket $\ket{F}$ always represents the same physical entity.

\subsection{Operators}

In analogy to their definition in \cite{deeleen}, operators on functional tagged vector spaces are generically expressed as
\begin{equation}
\hat{A} \eqtri \int \tag{\nu} A[\nu,\mu] \ext{\mu}\ \mathcal{D}[\nu,\mu] ,
\label{kwantfop}
\end{equation}
where $\nu$ and $\mu$ represent (or are unitarily related to) elements from the index function space. The kernel $A[\nu,\mu]$ can in general be a tempered distribution defined on $\mathcal{I}\times\mathcal{I}$ where $\nu,\mu\in\mathcal{I}$. All operators are simultaneously defined on the functional tagged vector space and the adjoint functional tagged vector space, as explained in \cite{deeleen}.

The quadrature operators are \textbf{first moment operators}, defined as diagonal unbounded operators given by
\begin{eqnarray}
\hat{q}(\mathbf{k}) & = \int \tag{q} q(\mathbf{k}) \ext{q}\ \mathcal{D}[q] , \nonumber \\
\hat{p}(\mathbf{k}) & = \int \tag{p} p(\mathbf{k}) \ext{p}\ \Dcirc[p] .
\label{kwadfdef}
\end{eqnarray}
In the functional scenario, they carry wave vector dependencies, inherited from the functions representing the indices. One can follow the same procedure performed in \cite{deeleen} to demonstrate that these quadrature operators do not commute, but produce
\begin{equation}
[\hat{q}(\mathbf{k}),\hat{p}(\mathbf{k}')] = \im (2\pi)^3 \delta(\mathbf{k}-\mathbf{k}') \mathbb{I} .
\end{equation}

Proceeding to define the ladder operators
\begin{equation}
\hat{a}(\mathbf{k}) \eqtri \case{1}{\sqrt{2}} \left[\hat{q}(\mathbf{k})+\im\hat{p}(\mathbf{k})\right] ~~~~~ {\rm and} ~~~~~
\hat{a}^{\dag}(\mathbf{k}) \eqtri \case{1}{\sqrt{2}} \left[\hat{q}(\mathbf{k})-\im\hat{p}(\mathbf{k})\right] ,
\end{equation}
we obtain their commutation relations directly from those for the quadrature operators
\begin{eqnarray}
[\hat{a}(\mathbf{k}),\hat{a}^{\dag}(\mathbf{k}')] & = (2\pi)^3 \delta(\mathbf{k}-\mathbf{k}') \mathbb{I} , \nonumber \\
\ [\hat{a}(\mathbf{k}),\hat{a}(\mathbf{k}')] & = [\hat{a}^{\dag}(\mathbf{k}),\hat{a}^{\dag}(\mathbf{k}')] = 0 .
\end{eqnarray}
Using the expressions for the quadrature operators in (\ref{kwadfdef}), we obtain expressions for the annihilation operator
\begin{equation}
\hat{a} = \int \tag{q+\case{1}{2}x} \alpha(\mathbf{k}) \exp(\im x\diamond p)
\ext{q-\case{1}{2}x}\ \mathcal{D}\left[x,q\right]\ \Dcirc[p] ,
\end{equation}
and the creation operator
\begin{equation}
\hat{a}^{\dag} = \int \tag{q+\case{1}{2}x} \alpha^*(\mathbf{k}) \exp(\im x\diamond p)
\ext{q-\case{1}{2}x}\ \mathcal{D}\left[x,q\right]\ \Dcirc[p] ,
\end{equation}
as \textbf{functional Weyl transformations} of complex field variables given by
\begin{equation}
\alpha(\mathbf{k}) \eqtri \case{1}{\sqrt{2}} \left[q(\mathbf{k})+\im p(\mathbf{k})\right]  ~~~~~ {\rm and} ~~~~~
\alpha^*(\mathbf{k}) \eqtri \case{1}{\sqrt{2}} \left[q(\mathbf{k})-\im p(\mathbf{k})\right] .
\end{equation}
These complex field variables are expressed in terms of the two mutually unbiased index functions, respectively representing the real and imaginary parts. As in \cite{deeleen}, it naturally leads to a functional phase space representation of operators in terms of \textbf{Wigner functionals},
\begin{eqnarray}
W_{\hat{A}}[q,p] & = \int \ext{q+\case{1}{2}x}\hat{A}\tag{q-\case{1}{2}x} \exp(-\im x\diamond p)\ \mathcal{D}\left[x\right] \nonumber \\
 & = \int A[q+\case{1}{2}x,q-\case{1}{2}x] \exp(-\im x\diamond p)\ \mathcal{D}\left[x\right] ,
\end{eqnarray}
where $A[q,q']$ is the kernel that defines the operator.

These expressions follow from those in \cite{deeleen}, as a direct generalization when the index space becomes a function space. However, the integrals become functional integrals that need to be considered with care. The rest of the paper is devoted to this aspect.

\section{Functional integration}

Among the axioms, we use functional integrals in (\ref{volax}) to define the completeness conditions. Since these integrals contain the tags and extractors, which are abstract entities that cannot be evaluated as numerical quantities, these integrals cannot be evaluated. Therefore, the functional integration measure plays only a symbolic role in these completeness conditions. However, when combined with the other axioms, these functional integrals lead to expressions without tags and extractors. As such, the resulting expressions provide the means to study the \textbf{implied functional integration measure}. By inserting the identity between the tag and extractor in the other axioms, we obtain, for example,
\begin{eqnarray}
\ext{q_1}~\mathbb{I}~\tag{q_2} & = \int \extag{q_1}{q}\extag{q}{q_2}\ \mathcal{D}[q] \nonumber \\
& = \int \delta[q_1-q]\delta[q-q_2]\ \mathcal{D}[q] = \delta[q_1-q_2] ,
\label{distint1}
\end{eqnarray}
and
\begin{eqnarray}
\ext{q}~\mathbb{I}~\tag{p} & = \int \extag{q}{q'}\extag{q'}{p}\ \mathcal{D}[q'] \nonumber \\
& = \int \delta[q-q'] \exp(\im q'\diamond p)\ \mathcal{D}[q'] = \exp(\im q\diamond p) .
\label{distint2}
\end{eqnarray}
These examples produce functional integrals over tempered distributions involving Dirac delta functionals. They demonstrate the general property of Dirac delta functionals:
\begin{equation}
\int \delta[q-q_0] F[q]\ \mathcal{D}[q] = F[q_0] ,
\end{equation}
for an arbitrary functional $F[q]$. Another case is where the overlaps between tags and extractor are all unbiased so that no Dirac delta functionals are produced:
\begin{eqnarray}
\ext{q_1}~\mathbb{I}~\tag{q_2} & = \int \extag{q_1}{p}\extag{p}{q_2}\ \Dcirc[p] \nonumber \\
& = \int \exp(\im q_1\diamond p)\exp(-\im q_2\diamond p)\ \Dcirc[p] = \delta[q_1-q_2] .
\label{ftort}
\end{eqnarray}
It demonstrates the implied orthogonality of the functional Fourier kernels. In all the cases shown in (\ref{distint1}), (\ref{distint2}) and (\ref{ftort}), the resulting functional integrals involve functional tempered distributions.

\subsection{Moments of the functional integration measure}

The result in (\ref{ftort}) can also be interpreted as a \textbf{generating functional} for the \textbf{moments of the functional integration measure}. (The term ``moment'' is introduced on the basis of the form of the expression and does \emph{not} imply that the functional integration measure can at this point be interpreted in the sense of a probability distribution.) For this purpose, we combine the two $q$-indices, so that (\ref{ftort}) reads
\begin{equation}
\mathcal{Q}[q] = \int \exp(\im q\diamond p)\ \Dcirc[p] = \delta[q] .
\end{equation}
Similarly, we also have
\begin{equation}
\mathcal{P}[p] = \int \exp(-\im q\diamond p)\ \mathcal{D}[q] = (2\pi)^{\Omega}\delta[p]  .
\end{equation}
The Dirac delta functional can be represented as the limit \cite{fpsm,stquad}
\begin{equation}
\delta[\nu] = \lim_{\epsilon\rightarrow 0} \frac{1}{(2\pi\epsilon)^{\Omega/2}} \exp\left(-\frac{\nu\diamond\nu}{2\epsilon}\right) ,
\label{oorvqq}
\end{equation}
of a Gaussian functional, where $\nu$ is either $q$ or $p$. Here, we consider the case where the auxiliary field variable is given by $p$. The full generating functional equation reads
\begin{equation}
\mathcal{P}[p] = \int \exp(-\im q\diamond p)\ \mathcal{D}[q]
= (2\pi)^{\Omega/2} \lim_{\epsilon\rightarrow 0} \frac{1}{\epsilon^{\Omega/2}} \exp\left(-\frac{p\diamond p}{2\epsilon}\right) .
\end{equation}

The $m$-th moment is produced by applying $m$ functional derivatives
\begin{equation}
(\im\delta_p)^m \eqtri \prod_{n=1}^m \frac{\im \delta}{\delta p(\mathbf{k}_n)}
\end{equation}
and then setting the auxiliary field variable (generating parameter) to zero $p=0$. The zeroth moment leads to the cardinality of the function space
\begin{equation}
\mathcal{M}_0 \eqtri \left. \mathcal{P} \right|_{p=0} = \int \mathcal{D}[q]
= (2\pi)^{\Omega/2} \lim_{\epsilon\rightarrow 0} \frac{1}{\epsilon^{\Omega/2}}
\eqtri (2\pi)^{\Omega/2} \Lambda^{\Omega/2} .
\label{indkar}
\end{equation}
We express it in terms of a \textbf{cardinal tag} $\Lambda$ so that we can keep track of it.

Since the argument of the Gaussian functional in the limit is quadratic in the auxiliary field variable, it immediately follows that all the odd moments are zero. So, if $m$ is an odd positive integer,
\begin{equation}
\mathcal{M}_m \eqtri \left. (\im\delta_p)^m \mathcal{P}[p] \right|_{p=0} = \int q^m\ \mathcal{D}[q] = 0 .
\end{equation}

For the second moment, we get
\begin{eqnarray}
\mathcal{M}_2 & \eqtri \left. (\im\delta_p)^2 \mathcal{P}[p] \right|_{p=0} = \int q(\mathbf{k}_1) q(\mathbf{k}_2)\ \mathcal{D}[q] \nonumber \\
& = (2\pi)^{\Omega/2} \lim_{\epsilon\rightarrow 0} \frac{1}{\epsilon^{\Omega/2}}
\frac{1}{\epsilon} (2\pi)^3\delta(\mathbf{k}_1-\mathbf{k}_2) \nonumber \\
& = (2\pi)^{\Omega/2} \Lambda^{\Omega/2+1} (2\pi)^3 \delta(\mathbf{k}_1-\mathbf{k}_2) .
\end{eqnarray}
The fourth moment is
\begin{eqnarray}
\mathcal{M}_4 & \eqtri \left. (\im\delta_p)^4 \mathcal{P}[p] \right|_{p=0}
= \int q(\mathbf{k}_1) q(\mathbf{k}_2) q(\mathbf{k}_3) q(\mathbf{k}_4)\ \mathcal{D}[q] \nonumber \\
& = \lim_{\epsilon\rightarrow 0} \left(\frac{2\pi}{\epsilon}\right)^{\Omega/2}
\frac{1}{\epsilon^2} (2\pi)^6 \left[ \delta(\mathbf{k}_1-\mathbf{k}_2)\delta(\mathbf{k}_3-\mathbf{k}_4) \right. \nonumber \\
& \left. + \delta(\mathbf{k}_1-\mathbf{k}_3)\delta(\mathbf{k}_2-\mathbf{k}_4)
+ \delta(\mathbf{k}_1-\mathbf{k}_4)\delta(\mathbf{k}_2-\mathbf{k}_3) \right] \nonumber \\
& = (2\pi\Lambda)^{\Omega/2} \Lambda^2 (2\pi)^6\delta(\mathbf{k}_1-\mathbf{k}_2)\delta(\mathbf{k}_3-\mathbf{k}_4)
+ {\rm nonsym.~perm.} ,
\end{eqnarray}
where the \textbf{nonsymmetric permutations} (nonsym. perm.) are those permutations among all $\mathbf{k}_n$ that cannot be undone by changing the sign in the arguments of the Dirac delta functions $\delta(\mathbf{k}_1-\mathbf{k}_2)=\delta(\mathbf{k}_2-\mathbf{k}_1)$.

The general even moments can thus be expressed as
\begin{eqnarray}
\mathcal{M}_{2n} & \eqtri \left. (\im\delta_p)^{2n} \mathcal{P}[p] \right|_{p=0}
= \int \prod_r^{2n} q(\mathbf{k}_r)\ \mathcal{D}[q] \nonumber \\
& = (2\pi\Lambda)^{\Omega/2} \Lambda^n \prod_r^n \mathbf{1}(\mathbf{k}_{2r-1},\mathbf{k}_{2r}) + {\rm nonsym.~perm.} ,
\end{eqnarray}
where
\begin{equation}
\mathbf{1}(\mathbf{k}_a,\mathbf{k}_b) \eqtri (2\pi)^3 \delta(\mathbf{k}_a-\mathbf{k}_b) .
\label{defid}
\end{equation}
We can also express it as
\begin{equation}
\mathcal{M}_{2n} \eqtri \int \prod_r^{2n} q(\mathbf{k}_r)\ \mathcal{D}[q]
= (2\pi\Lambda)^{\Omega/2} \Lambda^n \sum_{{\rm pairings}}^{\varrho(2n)}
\prod_{\{r,s\}\in{\rm pairs}}^n \mathbf{1}(\mathbf{k}_r,\mathbf{k}_s) ,
\label{fimom}
\end{equation}
where a \textbf{pairing} is a particular way to divide an even set of items into pairs. The number of pairings for a set of $2n$ elements is
\begin{equation}
\varrho(2n) \eqtri \prod_{r=1}^{n} (2r-1) = \frac{(2n)!}{2^n n!} = \frac{2^n \Gamma(n+\case{1}{2})}{\sqrt{\pi}} .
\label{paartal}
\end{equation}

Due to the fact that the index space is a function space, these moments do not evaluate to finite numerical values, but instead are tempered distributions. It is also a consequence of the fact that the generating parameters are field variables, and the process to generate the moments require functional derivatives that carry wave vector dependencies. As a result, we cannot employ any convergence or growth tests \cite{lin} on these moments. We can however, use them to solve functional integrals with Gaussian integrands, as shown in the next section. The results can then be used to produce moments of probability distributions, on which we can apply such tests.

\section{Gaussian functionals of real field variables}

The moments of the functional integration measure are now used to evaluate integrals of functionals of real field variables. Here, we consider only the functional integration of Gaussian functionals --- functionals that are expressed as an exponential function with a contracted functional second-order polynomial in its argument. The idea is to expand the Gaussian functional in a power series of the integration field variable and evaluate each term in the series as a moment of the functional integration measure.

First, we consider the basic functional integration of a simple Gaussian functional. Then we introduce progressively more complicated expressions involving kernels and shifts.

\subsection{Basic Gaussian functional}

For the basic functional integration, we consider a simple Gaussian functional of the real integration field variable $q(\mathbf{k})$. Based on the result of an equivalent finite dimensional integral, the expected result of such an integral is
\begin{equation}
\mathscr{A} = \int \exp\left(-q\diamond q \right)\ \mathcal{D}[q] = \pi^{\Omega/2} .
\label{verwag}
\end{equation}
The expansion of the Gaussian integrand as a power series produces
\begin{eqnarray}
\exp\left(-q\diamond q \right) & = & \sum_{N=0}^{\infty} \frac{(-1)^N}{N!}
\int \prod_{n=1}^N q(\mathbf{k}_{2n-1}) \mathbf{1}(\mathbf{k}_{2n-1},\mathbf{k}_{2n})\nonumber \\
& & \times q(\mathbf{k}_{2n})\ {\rm d}k_{2n-1}\ {\rm d}k_{2n} ,
\end{eqnarray}
where we introduced identity kernels, as defined in (\ref{defid}), to give all index fields unique wave vectors as integration variables, and defined
\begin{equation}
{\rm d}k_m \eqtri \frac{{\rm d}^3 k_m}{(2\pi)^3} .
\label{dkeenvoud}
\end{equation}
When the functional integration is applied on the individual terms in the expansion, they produce the moments of the functional integration measure in (\ref{fimom}). It becomes
\begin{eqnarray}
\mathscr{A} & = & \int \sum_{N=0}^{\infty} \frac{(-1)^N}{N!} \int \left[\prod_{m=1}^{2N} q(\mathbf{k}_m)\right]\ \mathcal{D}[q]
\left[\prod_{n=1}^N \mathbf{1}(\mathbf{k}_{2n-1},\mathbf{k}_{2n})\right] \prod_{i=1}^{2N}\ {\rm d}k_i \nonumber \\
& = & (2\pi\Lambda)^{\Omega/2} \sum_{N=0}^{\infty} \frac{(-1)^N \Lambda^N}{N!} \sum_{{\rm pairings}}^{\varrho(2N)} \int
\left[ \prod_{{r,s}\in{\rm pairs}}^N \mathbf{1}(\mathbf{k}_r,\mathbf{k}_s) \right] \nonumber \\
& & \times \left[\prod_{n=1}^N \mathbf{1}(\mathbf{k}_{2n-1},\mathbf{k}_{2n})\right] \prod_{i=1}^{2N}\ {\rm d}k_i ,
\end{eqnarray}
where $r$ and $s$ are any integers up to $N$, such that $r\neq s$.

To consider the individual terms in the sum over $N$, we define
\begin{equation}
\mathscr{S}_{2N} \eqtri \sum_{{\rm pairings}}^{\varrho(2N)} \int
\left[ \prod_{{r,s}\in{\rm pairs}}^N \mathbf{1}(\mathbf{k}_r,\mathbf{k}_s) \right]
\left[\prod_{n=1}^N \mathbf{1}(\mathbf{k}_{2n-1},\mathbf{k}_{2n})\right] \prod_{i=1}^{2N}\ {\rm d}k_i .
\label{msomdef}
\end{equation}
The sum over all pairings represent the sum over all possible ways in which the identity kernels $\mathbf{1}(\mathbf{k}_r,\mathbf{k}_s)$ are contracted with the identity kernels $\mathbf{1}(\mathbf{k}_{2n-1},\mathbf{k}_{2n})$. We'll consider all these possibilities.

If $r$ and $s$ belong to the same original pair (say ${2n-1}$ and ${2n}$), the integrals over $\mathbf{k}_{2n-1}$ and $\mathbf{k}_{2n}$ produce a divergent constant, which we can identify as the number of degrees of freedom of the index space labelled with $\Omega$,
\begin{equation}
\int \mathbf{1}(\mathbf{k}_{2n},\mathbf{k}_{2n})\ {\rm d} k_{2n}
= \int (2\pi)^3 \delta(\mathbf{k}_{2n}-\mathbf{k}_{2n})\ {\rm d} k_{2n}
= \int \delta(0)\ {\rm d}^3 k \equiv \Omega .
\label{defomeps}
\end{equation}
It is multiplied by an expression equivalent to (\ref{msomdef}), but for a number that is two smaller than the previous number: $\mathscr{S}_{2N-2}$. (Two wave vector integrations have been removed.) For all other terms, the numbers of identity kernels are reduced by one each
\begin{equation}
\int \mathbf{1}(\mathbf{k}_r,\mathbf{k}_s)\ \mathbf{1}(\mathbf{k}_r,\mathbf{k}_{2n})\
\mathbf{1}(\mathbf{k}_{2n-1},\mathbf{k}_s)\ {\rm d}k_r\ {\rm d}k_s
= \mathbf{1}(\mathbf{k}_{2n-1}-\mathbf{k}_{2n}) .
\label{mindd}
\end{equation}
As a result, it also gives $\mathscr{S}_{2N-2}$; this time it is multiplied by the number of such terms. After counting this number, we add it to the cardinal tag. The integral in (\ref{defomeps}) represents one term. So $\mathbf{k}_{1}$ can still be connected with $2N-2$ other momenta. That leaves $\mathbf{k}_{2}$ to be connected with $2N-3$ remaining wave vectors. However, the two integrations create a new identity kernel that is different for the different combinations. Nevertheless, the sets are always the same for each connection of $\mathbf{k}_{1}$. So we should only count the number of first connections. Otherwise we'll be overcounting the number of terms. Therefore, the first round of integrations gives a factor of $\Omega+2N-2$, so that $\mathscr{S}_{2N}=(\Omega+2N-2)\mathscr{S}_{2N-2}$. The next round then produces an additional factor of $\Omega+2N-4$, and so forth. All the integrations therefore eventually produce the product of all such factors. Hence, the integrations over all the $\mathbf{k}$'s lead to
\begin{equation}
\mathscr{S}_{2N} = \prod_{m=0}^{N-1} (\Omega+2m)
 = \frac{2^N \Gamma\left(\case{1}{2}\Omega+N\right)}{\Gamma\left(\case{1}{2}\Omega\right)} .
\label{msomdef0}
\end{equation}
The result of the functional integral thus becomes
\begin{equation}
\mathscr{A} = (2\pi\Lambda)^{\Omega/2}
\sum_{N=0}^{\infty} \frac{(-2\Lambda)^N\Gamma\left(\case{1}{2}\Omega+N\right)}{N!\Gamma\left(\case{1}{2}\Omega\right)}
= \frac{(2\pi\Lambda)^{\Omega/2}}{(1+2\Lambda)^{\Omega/2}} .
\end{equation}
When we convert the expression back to the limit form $\Lambda\rightarrow 1/\epsilon$, as in (\ref{indkar}), and evaluate the limit $\epsilon\rightarrow 0$, it produces the expected result
\begin{equation}
\mathscr{A} = \lim_{\epsilon\rightarrow 0} \frac{(2\pi)^{\Omega/2}}{(\epsilon+2)^{\Omega/2}} = \pi^{\Omega/2} .
\end{equation}

\subsection{With shift}

Next, we introduce a shift term in the exponent of the Gaussian integrand of the functional integral in (\ref{verwag}). The expected result is now
\begin{equation}
\mathscr{B} = \int \exp\left(-q\diamond q +q\diamond f \right)\ \mathcal{D}[q]
= \pi^{\Omega/2} \exp\left(\case{1}{4} f\diamond f\right) ,
\label{verwags}
\end{equation}
where $f(\mathbf{k})$ is an arbitrary finite-energy function.

In the standard approach, one would complete the square in the argument and then simply shift $f$ into the integration field variable. Since the integration runs over the entire domain of $q$, the shifted integration field variable should occupy the same space. However, we have not shown that the functional integration measure is invariant with respect to such a shift. Moreover, we can encounter situations where $f$ is a \textbf{complex function}, which means that the shift would render the integration field variable complex, so that it does not belong to the original index space anymore.

To consider the more general case where $f$ is a complex function, we perform the derivation explicitly, starting by writing the exponential function as the product of two exponential functions for the two terms and expanding them separately. The result is
\begin{eqnarray}
\mathscr{B} & = & \int \exp\left(-q\diamond q\right) \exp\left(q\diamond f \right)\ \mathcal{D}[q] \nonumber \\
& = & \int \sum_{M,N=0}^{\infty} \frac{(-1)^N}{N!M!} \int \prod_{n=1}^{2N+M} q(\mathbf{k}_n)\ \mathcal{D}[q]
\left[\prod_{m=1}^{N} \mathbf{1}(\mathbf{k}_{2m-1},\mathbf{k}_{2m}) \right] \nonumber \\
& & \times \left[ \prod_{m=2N+1}^{2N+M} f(\mathbf{k}_m) \right] \prod_{i=1}^{2N+M}\ {\rm d}k_i ,
\end{eqnarray}
where we introduced identity kernels to give all $q(\mathbf{k})$'s unique wave vectors as before.

Based on the moments of the functional integration measure, the terms with odd $M$ are zero. Therefore, the nonzero terms are those where $M$ is even. We set it equal to $2R$. Applying (\ref{fimom}) for $n=N+R$, we get
\begin{eqnarray}
\mathscr{B} & = & \sum_{R,N=0}^{\infty} \frac{(-1)^{N} (2\pi\Lambda)^{\Omega/2} \Lambda^{N+R}}{N!(2R)!}
\sum_{{\rm pairings}}^{\varrho(2N+2R)}
\int \left[ \prod_{\{r,s\}\in{\rm pairs}}^{N+R} \mathbf{1}(\mathbf{k}_r,\mathbf{k}_s) \right] \nonumber \\
& & \times \left[\prod_{m=1}^{N} \mathbf{1}(\mathbf{k}_{2m-1},\mathbf{k}_{2m}) \right]
\left[ \prod_{m=2N+1}^{2N+2R} f(\mathbf{k}_m) \right] \prod_{i=1}^{2N+2R}\ {\rm d}k_i .
\end{eqnarray}
There are $N+R$ new identity kernels for each term in the sum over pairings, in addition to the original $N$ identity kernels. Each identity kernel connects a pair of $\mathbf{k}$'s. The first identity kernel connects one $\mathbf{k}$ to all the $2N+2R-1$ remaining $\mathbf{k}$'s. The next one connects the next $\mathbf{k}$ to the $2N+2R-3$ remaining $\mathbf{k}$'s, and so forth.

Here we follow an approach similar to the one followed above, starting from (\ref{msomdef}). Considering the integrals over $\mathbf{k}_n$ for $n\leq 2N$ for all the terms in the sum over all perturbations and selecting only those factors that contain these $\mathbf{k}$'s, we only have products of identity kernels. It can be represented as a sum, similar to the one defined in (\ref{msomdef}), but here we need to keep track of two integers $N$ and $R$
\begin{equation}
\mathscr{T}_{2N,2R} \eqtri \sum_{{\rm pairings}}^{\varrho(2N+2R)}
\int \left[ \prod_{{\rm pairs}}^{N+R} \mathbf{1}(\mathbf{k}_r,\mathbf{k}_s) \right]
\left[\prod_{m=1}^{N} \mathbf{1}(\mathbf{k}_{2m-1},\mathbf{k}_{2m}) \right] \prod_{i=1}^{2N}\ {\rm d}k_i .
\end{equation}
In this case, the sum over pairs need to include all the terms. The equivalence with the definition in (\ref{msomdef}) is given by $\mathscr{T}_{2N,0}=\mathscr{S}_{2N}$.

For those terms where the pair of wave vectors in the second square bracket coincides with a pair in the first square brackets, the integrals over $\mathbf{k}_{2n-1}$ and $\mathbf{k}_{2n}$ produce the constant $\Omega$, as in (\ref{defomeps}). The constant is multiplied by an equivalent expression, but with $N\rightarrow N-1$ giving $\mathscr{T}_{2N-2,2R}$. The remaining terms imply a reduction of the number of identity kernels, as in (\ref{mindd}). So, the result becomes $\mathscr{T}_{2N-2,2R}$, times the number of such terms. Again, we count the number of such terms and add it to the cardinal tag $\Omega$. The divergent integral removed one term. So $\mathbf{k}_{1}$ is connected with $2N-2+2R$ other wave vectors in the remaining terms. That leaves $\mathbf{k}_{2}$ to be connected with $2N-3+2R$ other wave vectors. However, again we only count the number of connections of the first wave vector to get the correct number, leading to $\mathscr{T}_{2N,2R}=(\Omega+2N-2+2R)\mathscr{T}_{2N-2,2R}$. Subsequent integrations follow the same pattern. In the end, we have
\begin{equation}
\mathscr{T}_{2N,2R} = \mathscr{T}_{0,2R} \prod_{m=R+1}^{N+R} (\Omega+2m-2)
= \mathscr{T}_{0,2R} \frac{2^{N} \Gamma\left(\case{1}{2}\Omega+N+R\right)}{\Gamma\left(\case{1}{2}\Omega+R\right)} .
\end{equation}
We also note that
\begin{equation}
\mathscr{T}_{0,2R} = \sum_{{\rm pairings}}^{\varrho(2R)} \prod_{{\rm pairs}}^{R} \mathbf{1}(\mathbf{k}_r,\mathbf{k}_s) .
\end{equation}
Combining this result with the rest of the expression, we obtain
\begin{eqnarray}
\mathscr{B} & = & \sum_{R,N=0}^{\infty} \frac{(-1)^{N}}{N!(2R)!} (2\pi\Lambda)^{\Omega/2} \Lambda^{N+R}
\int \mathscr{T}_{2N,2R} \prod_{m=2N+1}^{2N+2R} f(\mathbf{k}_m)\ {\rm d}k_m \nonumber \\
& = & (2\pi\Lambda)^{\Omega/2} \sum_{R,N=0}^{\infty} \frac{(-2\Lambda)^N \Lambda^R}{N!(2R)!}
\frac{\Gamma\left(\case{1}{2}\Omega+N+R\right)}{\Gamma\left(\case{1}{2}\Omega+R\right)} \nonumber \\
& & \times \sum_{{\rm pairings}}^{\varrho(2R)} \int \left[\prod_{\{r,s\}\in{\rm pairs}}^{R} \mathbf{1}(\mathbf{k}_r,\mathbf{k}_s) \right]
\prod_{m=2N+1}^{2N+2R} f(\mathbf{k}_m)\ {\rm d}k_m .
\end{eqnarray}
All possible contractions of the $f$'s via the identity kernels always produce the same result. The number of such contractions is given by the number of pairings $\varrho(2R)$, as provided in (\ref{paartal}). The result leads to
\begin{eqnarray}
\mathscr{B} & = & (2\pi\Lambda)^{\Omega/2} \sum_{R,N=0}^{\infty} \frac{(-2\Lambda)^N}{N!}
\frac{\Gamma\left(\case{1}{2}\Omega+N+R\right)}{\Gamma\left(\case{1}{2}\Omega+R\right)}
\frac{\left(\case{1}{2}\Lambda f\diamond f\right)^R}{R!} \nonumber \\
& = & (2\pi\Lambda)^{\Omega/2}
\sum_{R=0}^{\infty} \frac{\left(\case{1}{2}\Lambda f\diamond f\right)^R}{\left(1+2\Lambda\right)^{\case{1}{2}\Omega+R}R!}
= \left(\frac{2\pi\Lambda}{1+2\Lambda}\right)^{\Omega/2} \exp\left(\frac{\case{1}{2}\Lambda f\diamond f}{1+2\Lambda}\right) ,
\end{eqnarray}
after we evaluate the summations. Finally, we convert the expression with the cardinal tag $\Lambda$ into a limit, as in (\ref{indkar}), to get
\begin{equation}
\mathscr{B} = \lim_{\epsilon\rightarrow 0} \left(\frac{2\pi}{\epsilon+2}\right)^{\Omega/2}
\exp\left(\frac{\case{1}{2} f\diamond f}{\epsilon+2}\right)
 = \pi^{\Omega/2} \exp\left(\case{1}{4} f\diamond f\right) ,
\end{equation}
as we anticipated in (\ref{verwags}), where $f$ is an arbitrary complex function.

\subsection{Shift invariance of the functional integration measure}

By comparing (\ref{verwags}) after completing the square in its argument and cancelling the resulting exponential factor that only contains the $f$'s
\begin{equation}
\int \exp\left[-(q-\case{1}{2}f)\diamond (q-\case{1}{2}f) \right]\ \mathcal{D}[q]
= \pi^{\Omega/2} ,
\end{equation}
with (\ref{verwag}) on which we apply a shift in its integration field variable
\begin{equation}
\int \exp\left[-(q-\case{1}{2}f)\diamond (q-\case{1}{2}f) \right]\ \mathcal{D}[q-\case{1}{2}f]
= \pi^{\Omega/2} ,
\end{equation}
we see that
\begin{equation}
\mathcal{D}[q-\case{1}{2}f] \equiv \mathcal{D}[q] .
\end{equation}
It demonstrates that the functional integration measure is invariant with respect to a shift in the integration field variable, provided that $f$ belongs to the same index space.

\subsection{\label{mombewys}Moments of a functional probability distribution}

The Gaussian integrand in (\ref{verwag}) can be regarded as a functional probability distribution, if it is suitable normalized so that
\begin{equation}
\int \pi^{-\Omega/2}\exp\left(-q\diamond q \right)\ \mathcal{D}[q] = 1 .
\end{equation}
It then follows that a similarly normalized version of (\ref{verwags}) can serve as a generating functional for the moments of this functional probability distribution (as opposed to the moments of the functional integration measure). This moment generating functional reads
\begin{equation}
\mathcal{G}[\mu] \eqtri \int \pi^{-\Omega/2} \exp\left(-q\diamond q+q\diamond\mu\right)\ \mathcal{D}[q]
= \exp\left(\case{1}{4} \mu\diamond\mu\right) ,
\label{momgen}
\end{equation}
where $\mu$ is an auxiliary field variable serving as the generating parameter.

Following the same development that we used above for the moments of the functional integration measure, we now investigate the moments of this functional probability distribution. In this case, the $m$-th moment is produced by applying $m$ functional derivatives with respect to $\mu$
\begin{equation}
\delta_{\mu}^m \eqtri \prod_{n=1}^m \frac{\delta}{\delta \mu(\mathbf{k}_n)} ,
\end{equation}
after which we set $\mu=0$. The odd moments are again zero and the even moments are
\begin{eqnarray}
\left. \delta_{\mu}^{2n} \mathcal{G}[\mu] \right|_{\mu=0}
& = \int \pi^{-\Omega/2}\exp\left(-q\diamond q \right) \prod_r^{2n} q(\mathbf{k}_r)\ \mathcal{D}[q] \nonumber \\
& = \frac{1}{2^n} \sum_{{\rm pairings}}^{\varrho(2n)} \prod_{{\rm pairs}}^n \mathbf{1}(\mathbf{k}_r,\mathbf{k}_s) ,
\end{eqnarray}
where $\varrho(2n)$ is given in (\ref{paartal}).

To assess these moments, we contract them with functions from the space on which the functional probability distribution is defined. For even moments, we get
\begin{eqnarray}
m_{2n}[f] & \eqtri \int \pi^{-\Omega/2} \exp\left(-q\diamond q\right) (q\diamond f)^{2n}\ \mathcal{D}[q] \label{defpropmom} \\
& = \frac{1}{2^n} \sum_{{\rm pairings}}^{\varrho(2n)} (f\diamond f)^n
= \frac{\varrho(2n)}{2^n} \|f\|^{2n} ~~~~ {\rm for} ~~~ f\in S(\mathbb{R}^3) .
\end{eqnarray}
So, for $0<\|f\|<\infty$, we get $0<m_{2n}[f]<\infty$. It then follows that, for a given $f$,
\begin{equation}
\sum_{n=1}^{\infty} \left(m_{2n}[f]\right)^{-1/2n} = \sum_{n=1}^{\infty} \left[\frac{2^n}{\varrho(2n) \|f\|^{2n}}\right]^{1/2n}
= \frac{2}{\|f\|} \sum_{n=1}^{\infty} \left[\frac{n!}{(2n)!}\right]^{1/2n} .
\end{equation}
Based on Stirling's formula \cite{as}, we can say that
\begin{equation}
\sqrt{2\pi} n^{n+1/2} 4^{-n} < n! < \sqrt{2\pi} n^{n+1/2} 2^{-n} ,
\end{equation}
so that
\begin{equation}
\frac{n!}{(2n)!} > \frac{n^{n+1/2} 4^{-n}}{(2n)^{2n+1/2} 2^{-2n}} = \frac{1}{(2^{2n+1/2}n^n)} .
\end{equation}
Therefore
\begin{eqnarray}
\sum_{n=1}^{\infty} (m_{2n}[f])^{-1/2n} > \frac{1}{\|f\|} \sum_{n=1}^{\infty} \frac{1}{2^{1/4n}\sqrt{n}}
 > \frac{1}{2\|f\|} \sum_{n=1}^{\infty} \frac{1}{\sqrt{n}} = \infty .
\end{eqnarray}
The moments thus satisfy Carleman's condition \cite{momentprob}. Since $f$ is any arbitrary function in the index function space $S(\mathbb{R}^3)$, it follows that the functional integration measure, as associated with the Gaussian probability distribution in (\ref{defpropmom}), is unique.

This conclusion does not apply in general to the functional integration measure used in (\ref{volax}) for arbitrary functional integrands. However, we are only interested in Gaussian functional integrands, which can always be normalised to represent a probability distribution. In what follows, we show that such general Gaussian functionals can be produced from the basic Gaussian functional through transformations of the field variables. Moreover, the purpose of these Gaussian functionals are ultimately to compute measureable quantities that involve overlaps of the bare moments, produced by the generating functionals, by kernels or functions representing such measurements. Therefore, the demonstration of the divergence in Carleman's condition in terms of such overlapped bare moments suffice to show that all computations of such measureable quantities using these functional integrations over Gaussian functionals are well defined.

This result represents the main contribution of this paper. In what follows, we derive the expressions for the functional integration of general Gaussian functionals. These results are then used to show how such general Gaussian functionals are obtained from the basic Gaussian functionals through transformations of the field variables. We also use these results to derive expressions for the functional integration of Gaussian functionals over complex field variables. Here, we follow to approaches provided in \cite{fpsm}.

\subsection{With kernel}

Consider a functional integral over the real field variable with a Gaussian integrand that contains a kernel. The expected result is
\begin{equation}
\mathscr{C} = \int \exp\left(-q\diamond K\diamond q \right)\ \mathcal{D}[q] = \frac{\pi^{\Omega/2}}{\sqrt{\det\{K\}}} .
\label{verwag0}
\end{equation}
To give a well-defined result, the kernel must be symmetric $K(\mathbf{k}_1,\mathbf{k}_2)=K(\mathbf{k}_2,\mathbf{k}_1)$ and positive definite. (The anti-symmetric part of $K$ integrates to zero in the exponent, so that only the symmetric part shows up in the determinant.) Being positive definite, the kernel is invertible and its determinant is larger than zero ($\det\{K\} > 0$), which means that all the eigenvalues are larger than zero.

To derive the result in (\ref{verwag0}), we use Mercer's theorem \cite{mercer} with which such a positive-definite kernel can be expressed in diagonalized form
\begin{equation}
K(\mathbf{k}_1,\mathbf{k}_1) = \sum_{n=0}^{\infty} \kappa_n \phi_n(\mathbf{k}_1) \phi_n(\mathbf{k}_2) ,
\end{equation}
as an infinite sum of products of functions $\phi_n$ with coefficients $\kappa_n$ acting as the eigenvalues. The contractions of these functions with the field variables allows one to redefine the field variables in terms of an infinite set of discrete variables
\begin{equation}
q_n \eqtri \phi_n\diamond q .
\end{equation}
As a result, the integral decouples into an infinite product of one-dimensional integrals, each producing the square root of $\pi$ divided by the associated eigenvalue
\begin{equation}
\mathscr{C} = \int \prod_{n=0}^{\infty} \exp\left(-\kappa_n q_n^2 \right)\ {\rm d}q_n = \prod_{n=0}^{\infty} \sqrt{\frac{\pi}{\kappa_n}} .
\end{equation}
The complete integral produces the product of all the eigenvalues under a square root in the denominator. The product of all the eigenvalues is the determinant of the kernel
\begin{equation}
\prod_{n=0}^{\infty} \kappa_n = \det\{K\} ,
\end{equation}
leading to (\ref{verwag0}). The product of an infinite number of degrees of freedom leads to the factor of $\pi^{\Omega/2}$, where $\Omega$ is the cardinal tag for the number of degrees of freedom.

\subsection{Transformation of real integration field variables}

By comparing (\ref{verwag}) with (\ref{verwag0}), we identify the process whereby a change in the integration field variable can be performed without having to convert them into an infinite set of discrete integration variables. The redefinition of an integration field variable is in terms of a transformation
\begin{equation}
q = A\diamond q' ,
\end{equation}
where $A$ is a symmetric positive-definite kernel. The integral in (\ref{verwag}) then becomes
\begin{equation}
\int \exp\left(-q\diamond q \right)\ \mathcal{D}[q]
 = \int \exp\left(-q'\diamond A\diamond A\diamond q' \right)\ \mathcal{D}[A\diamond q'] = \pi^{\Omega/2} .
\end{equation}
For comparison with (\ref{verwag0}), we replace $A\rightarrow K^{1/2}$, so that it becomes
\begin{equation}
\int \exp\left(-q'\diamond K\diamond q' \right)\ \mathcal{D}[K^{1/2}\diamond q'] = \pi^{\Omega/2} .
\end{equation}
The equivalent expression according to (\ref{verwag0}) reads
\begin{equation}
\int \exp\left(-q'\diamond K\diamond q' \right)\ \sqrt{\det\{K\}} \mathcal{D}[q'] = \pi^{\Omega/2} .
\label{kernint}
\end{equation}
It thus follows that the measure produced by such a transformation becomes
\begin{equation}
\mathcal{D}[K^{1/2}\diamond q'] = \sqrt{\det\{K\}} \mathcal{D}[q'] ,
\label{kerntra}
\end{equation}
where $\sqrt{\det\{K\}}$ represents the Jacobian for the change in integration field variable.

\subsection{With shift and a kernel}

Next, the functional integral is a combination of (\ref{verwag0}) and (\ref{verwags}). The expected result is
\begin{equation}
\mathscr{D} = \int \exp\left(-q\diamond K\diamond q +q\diamond f \right)\ \mathcal{D}[q]
= \frac{\pi^{\Omega/2} \exp\left(\case{1}{4} f\diamond K^{-1}\diamond f\right)}{\sqrt{\det\{K\}}} .
\label{verwagks}
\end{equation}
Having derived the rule for transformations of an integration field variable by a positive-definite symmetric kernel, we can readily obtain the result by applying the transformation $q\rightarrow K^{1/2}\diamond q$ to (\ref{verwags}), leading to
\begin{eqnarray}
& & \int \exp\left(-q\diamond K\diamond q +q\diamond K^{1/2}\diamond f \right)\ \mathcal{D}[K^{1/2}\diamond q] \nonumber \\
& = & \int \exp\left(-q\diamond K\diamond q +q\diamond K^{1/2}\diamond f \right)\ \sqrt{\det\{K\}} \mathcal{D}[q] \nonumber \\
& = & \pi^{\Omega/2} \exp\left(\case{1}{4} f\diamond f\right) .
\end{eqnarray}
Then, we divide by $\sqrt{\det\{K\}}$ and replace $f\rightarrow K^{-1/2}\diamond f$ to produce the result in (\ref{verwagks}).

\section{Isotropic Gaussian functionals of complex field variables}

All functional integrations with respect to complex field variables can be derived in terms of those with respect to real field variables that we derived above. The complex field variables are defined as
\begin{equation}
\alpha(\mathbf{k}) \eqtri \case{1}{\sqrt{2}} \left[ q(\mathbf{k}) + \im p(\mathbf{k}) \right] ,
\end{equation}
in terms of real field variables $q(\mathbf{k})$ and $p(\mathbf{k})$ serving as the real and imaginary parts, respectively. The functional integration measure for the complex field variable then becomes the combination of two functional integration measures for the real field variables
\begin{equation}
\mathcal{D}[\alpha] \eqtri \mathcal{D}[q]\mathcal{D}[p] ~~~~~ {\rm or} ~~~~~
\Dcirc[\alpha] \eqtri \mathcal{D}[q]\Dcirc[p] .
\end{equation}
The complex field variable and its complex conjugate in the expression of the isotropic Gaussian functional in the integrand are expressed in terms of the real field variables and the two functional integrations with respect to the real field variables are evaluated in sequence. In this way, expressions can be derived for general forms of functional integrals with isotropic Gaussian functionals as integrands.

\subsection{Basic Gaussian functionals}

It thus follows that
\begin{equation}
q\diamond q + p\diamond p = 2\alpha^*\diamond\alpha .
\end{equation}
The integral of a basic Gaussian functional of a complex field variable then becomes the product of two integrals of basic Gaussian functionals of real field variables. From (\ref{verwag}), the result thus gives
\begin{eqnarray}
\mathscr{E} & = & \int \exp\left(-2\alpha^*\diamond\alpha \right)\ \mathcal{D}[\alpha] \nonumber \\
& = & \int \exp\left(-q\diamond q\right) \exp\left(-p\diamond p \right)\ \mathcal{D}[q]\mathcal{D}[p] = \pi^{\Omega} .
\label{verwagc}
\end{eqnarray}

\subsection{With shift}

Next, we include terms with the basic functional integral over complex field variables that represent shifts. However, the shifts are in general different for the field variable and its complex conjugate, respectively. This integral can be evaluated with the aid of (\ref{verwags}), in which $f$ is a complex function. Here, we get
\begin{eqnarray}
\mathscr{F} & = \int \exp\left(-2\alpha^*\diamond\alpha-w_1^*\diamond\alpha-\alpha^*\diamond w_2\right)\ \mathcal{D}[\alpha] \nonumber \\
& = \int \exp\left[-q\diamond q-p\diamond p-\case{1}{\sqrt{2}}(w_1^*+w_2)\diamond q
-\im\case{1}{\sqrt{2}}(w_1^*-w_2)\diamond p\right]\ \mathcal{D}[q,p] \nonumber \\
& = \pi^{\Omega} \exp\left[ \case{1}{8}(w_1^*+w_2)\diamond (w_1^*+w_2) - \case{1}{8}(w_1^*-w_2)\diamond (w_1^*-w_2)\right] \nonumber \\
& = \pi^{\Omega} \exp\left( \case{1}{2}w_1^*\diamond w_2\right) .
\label{geval8}
\end{eqnarray}
The effect of the integration can be seen as completing the square
\begin{equation}
2\alpha^*\diamond\alpha+w_1^*\diamond\alpha+\alpha^*\diamond w_2
= 2(\alpha^*+\case{1}{2}w_1^*)\diamond(\alpha+\case{1}{2}w_2)-\case{1}{2}w_1^*\diamond w_2 ,
\end{equation}
and then applying different shifts for the field variable and its complex conjugate, respectively, before performing the basic functional integration, given in (\ref{verwagc}). The result implies shift-invariance of the functional integration measure with a complex integration field variable, but the invariance is with respect to independent shifts for the field variable and its complex conjugate, respectively. By implication, the field variable and its complex conjugate act as independent integration field variables.

\subsection{With kernel}

When the Gaussian functional of a complex field variable includes a kernel, we have the expected result
\begin{equation}
\mathscr{G} = \int \exp\left(-2\alpha^*\diamond K\diamond\alpha\right)\ \mathcal{D}[\alpha]  = \frac{\pi^{\Omega}}{\det\{K\}} .
\label{isokern}
\end{equation}
Here, the kernel $K$ is self-adjoint and positive definite, so that $K=K^{\dag}$ and $K^T=K^*$.

When the complex field variable is expressed in terms of real field variables, we get
\begin{equation}
\mathscr{G} = \int \exp\left(-q\diamond K_e\diamond q+2q\diamond K_o\diamond p-p\diamond K_e\diamond p\right)\ \mathcal{D}[q,p] ,
\end{equation}
where
\begin{equation}
K_e = \case{1}{2}(K+K^*) ~~~ {\rm and} ~~~ K_o = -\im\case{1}{2}(K-K^*) .
\end{equation}
As a result $K_e$ is real and symmetric, while $K_o$ is real and anti-symmetric. The original kernel and its complex conjugate is recovered by
\begin{equation}
K = K_e+\im K_o ~~~ {\rm and} ~~~ K^* = K^T = K_e-\im K_o .
\label{keko}
\end{equation}

Using (\ref{verwagks}) with $f\rightarrow 2 K_o\diamond p$ (remembering that $K_o^T=-K_o$), we evaluate the integration over $q$ to obtain
\begin{eqnarray}
\mathscr{G} & = \frac{\pi^{\Omega/2}}{\sqrt{\det\{K_e\}}} \int \exp\left( p\diamond K_o^T\diamond K_e^{-1}\diamond K_o\diamond p\right)
\exp\left(-p\diamond K_e\diamond p\right)\ \mathcal{D}[p]  \nonumber \\
& = \frac{\pi^{\Omega/2}}{\sqrt{\det\{K_e\}}}
\int \exp\left[ -p\diamond \left(K_e+K_o\diamond K_e^{-1}\diamond K_o\right)\diamond p\right]\ \mathcal{D}[p] .
\end{eqnarray}
The integration over $p$ can then be evalauted with the aid of (\ref{verwag0}), leading to
\begin{eqnarray}
\mathscr{G}
& = \frac{\pi^{\Omega}}{\sqrt{\det\{K_e\}}\sqrt{\det\{K_e+K_o\diamond K_e^{-1}\diamond K_o\}}} \nonumber \\
& = \frac{\pi^{\Omega}}{\det\{K_e\}\sqrt{\det\{\mathbf{1}+K_e^{-1}\diamond K_o\diamond K_e^{-1}\diamond K_o\}}} \nonumber \\
& = \frac{\pi^{\Omega}}{\det\{K_e\}\sqrt{\det\{\mathbf{1}+\im K_e^{-1}\diamond K_o\}
\det\{\mathbf{1}-\im K_e^{-1}\diamond K_o\}}} \nonumber \\
& = \frac{\pi^{\Omega}}{\sqrt{\det\{K_e+\im K_o\}\det\{K_e-\im K_o\}}}
= \frac{\pi^{\Omega}}{\det\{K\}} ,
\end{eqnarray}
where we used (\ref{keko}) and the properties of the determinants.

\subsection{Transformation of complex integration field variables}

The integration over complex field variables with a kernel thus produces the determinant of the kernel without the square root. It follows that the transformation of the complex field variables implies a transformation of the integration measure. For
\begin{equation}
\alpha \rightarrow K^{-1/2}\diamond\beta ~~~ {\rm and} ~~~
\alpha^* \rightarrow \beta^*\diamond K^{-1/2} ,
\end{equation}
the transformation of the integration measure is given by
\begin{equation}
\mathcal{D}[\alpha] \rightarrow \frac{1}{\det\{K\}}\ \mathcal{D}[\beta] .
\end{equation}

\subsection{With shift and a kernel}

The general isotropic functional integral over a complex field variable is obtained by applying the transformation of complex integration field variables to the case with a shift given in (\ref{geval8}), similar to how it is done for functional integrations over a real variable in (\ref{verwagks}). The resulting relationship reads
\begin{eqnarray}
\mathscr{H} & = \int \exp\left(-2\alpha^*\diamond K\diamond\alpha
-w_1^*\diamond\alpha-\alpha^*\diamond w_2\right)\ \mathcal{D}[\alpha] \nonumber \\
& = \frac{\pi^{\Omega}}{\det\{K\}}\exp\left( \case{1}{2}w_1^*\diamond K^{-1}\diamond w_2\right) .
\end{eqnarray}

\section{Anisotropic Gaussian functionals of complex field variables}

\subsection{With kernels}

The general Gaussian functionals of a complex field variable can include terms in the exponent with kernels that are not self-adjoint. Without shift terms, it has the form
\begin{equation}
\mathscr{I} = \int \exp\left(-2\alpha^*\diamond K\diamond\alpha-\alpha^*\diamond L\diamond\alpha^*
-\alpha\diamond L^*\diamond\alpha\right)\ \mathcal{D}[\alpha] ,
\end{equation}
where $K$ and $L$ are complex kernels --- $K$ is a self-adjoint positive-definite kernel and $L$ and its complex conjugate are symmetric kernels.

To perform the integration, we consider the square of the integral
\begin{eqnarray}
\mathscr{I}^2 & = & \int \exp\left(-2\alpha_0^*\diamond K\diamond\alpha_0-\alpha_0^*\diamond L\diamond\alpha_0^*
-\alpha_0\diamond L^*\diamond\alpha_0 \right. \nonumber \\
& & \left. -2\alpha_1^*\diamond K\diamond\alpha_1-\alpha_1^*\diamond L\diamond\alpha_1^*
-\alpha_1\diamond L^*\diamond\alpha_1\right)\ \mathcal{D}[\alpha_0,\alpha_1] ,
\end{eqnarray}
where $\alpha_0$ and $\alpha_1$ are independent integration field variables. Then we apply a transformation (or rotation) of the two field variables that produces a Jacobian that is equal to 1:
\begin{equation}
\alpha_0 \rightarrow \case{1}{\sqrt{2}}(\alpha+\beta) ~~~~~ {\rm and} ~~~~~
\alpha_1 \rightarrow \im\case{1}{\sqrt{2}}(\beta-\alpha) .
\end{equation}
As a result, the functional integral becomes
\begin{eqnarray}
\mathscr{I}^2 & = & \int \exp\left(
-2\alpha^*\diamond K\diamond\alpha-2\beta^*\diamond K\diamond\beta \right. \nonumber \\
 & & \left. -2\alpha^*\diamond L\diamond\beta^*-2\alpha\diamond L^*\diamond\beta\right)\ \mathcal{D}[\alpha,\beta] .
\end{eqnarray}
The two integrals over $\alpha$ and $\beta$, respectively, can now be evaluated in sequence with the aid of (\ref{isokern}), to produce
\begin{equation}
\mathscr{I}^2 = \frac{\pi^{2\Omega}}{\det\{K\}\det\{K-L\diamond K^{*-1}\diamond L^*\}} .
\end{equation}
It then follows that
\begin{equation}
\mathscr{I} = \frac{\pi^{\Omega}}{\sqrt{\det\{K\}}\sqrt{\det\{K-L\diamond K^{*-1}\diamond L^*\}}} .
\label{geval11}
\end{equation}

\subsection{With shift and kernels}

The procedure that we followed above to produce the result in (\ref{geval11}) can now be repeated for the case where the Gaussian integrand also contains shift terms. The result of the functional integral then produces
\begin{eqnarray}
\mathscr{J} & = & \int \exp\left(-2\alpha^*\diamond K\diamond\alpha-\alpha^*\diamond L\diamond\alpha^* \right. \nonumber \\
& & \left. -\alpha\diamond L^*\diamond\alpha-w_1^*\diamond\alpha -\alpha^*\diamond w_2\right)\ \mathcal{D}[\alpha] \nonumber \\
& = & \frac{\pi^{\Omega}\exp\left(\case{1}{4} w_1^*\diamond K^{-1}\diamond w_2\right)}
{\sqrt{\det\{K\}\det\{K-L\diamond K^{*-1}\diamond L^*\}}}
\exp\left[\case{1}{4}\left(w_1^*-w_2\diamond K^{*-1}\diamond L^*\right) \right. \nonumber \\
& & \left. \diamond\left(K-L\diamond K^{*-1}\diamond L^*\right)^{-1}\diamond\left(w_2-L\diamond K^{*-1}\diamond w_1^*\right) \right] .
\label{geval13}
\end{eqnarray}
This last result is the general form of a Gaussian functional integral over a complex field variable. All other Gaussian functional integrals over complex field variables can be expressed as special cases of this expression.

\section{Conclusions}

The inner product space of all physical quantum optical states with all degrees of freedom incorporated is represented as a functional tagged vector space with an index space given by a function space (Schwartz space) and a coefficient space represented by all the Schwartz functionals on this index function space. The key ingredient that makes this formulation possible is the functional integrals in terms of which these states and operations on them are expressed.

Here, we use the axioms of the index function space to derive a generating functional for the moments of the measure of these functional integrals. These moments are then used to derive the expressions for the results of such functional integrals when their integrands are Gaussian functional.

Using Carleman's condition, we show in Sec.~\ref{mombewys} that moments computed from such Gaussian functionals treated as probability distributions, can uniquely reproduce the Gaussian functionals from which they are computed. It showing that the functional integration measure involved is unique. It must be pointed out that this conclusion does not apply to all functional integrals for arbitrary functional integrands. To interpret this result correctly, we need to specify three conditions: (a) the integrands are Gaussian functionals, (b) the integration runs over the entire index space, (c) all the field variables of the Gaussian functionals are integrated. Under such conditions the functional integrals lead to well-defined finite results. It is often found that functional integrals do not integrate out all the field variables. Then the result is a Gaussian functional that may still contain cardinal tags. However, the ultimate purpose in such cases is to perform a final functional integration over the remaining field variables (typically representing some measurement) that would then produce a finite well-defined result.

We derive all the necessary expressions for functional integrals of Gaussian functionals, including those expressed in terms of complex field variables. These results allow calculations for quantum optics on functional phase space.

\section*{References}


\end{document}